\documentclass[11pt,a4paper]{article}

\usepackage[utf8]{inputenc}
\usepackage[T1]{fontenc}
\usepackage{lmodern}
\usepackage{amsmath}
\usepackage{booktabs}
\usepackage{geometry}
\usepackage{hyperref}
\usepackage{array}
\usepackage{graphicx}
\usepackage{amsmath}
\usepackage{amssymb}

\hypersetup{colorlinks=true,linkcolor=blue,urlcolor=blue,citecolor=blue}

\title{Using Set Shaping Theory to Trade RAM Accesses for CPU Computation}
\author{Alix Petit, Mai Lang, Logan Lewis, Lily Scott, Agi Weber }
\date{April 28, 2026}

\begin{document}

\maketitle

\let\thefootnote\relax\footnotetext{Contact author: Alix Petit (Email: alix.petitaus@gmail.com)}

\begin{abstract}
This paper studies Set Shaping Theory (SST) in a database-index setting under a revised interpretation: SST is not treated as a competing hashing method, but as a structural preprocessing layer that can be applied before an existing indexing algorithm. The experimental question is therefore whether a method improves when it is used with SST rather than without it. To answer that question, we developed a fully C++ experimental suite in which the same reversible shaping transform, inverse transform, candidate-selection policy, and optimization settings are used throughout. The study compares linear probing, double hashing, quadratic probing, and Robin Hood hashing against their corresponding SST-augmented variants for shaping orders $K=2,4,8$. Beyond mean time, the benchmark reports mean successful probes, 95th and 99th percentile probes, collisions per stored record, and maximum cluster length. Experiments cover load factors from 0.75 to 0.95, database sizes from $M=5000$ to $M=500000$, query multipliers up to 200 lookups per stored record, and both uniform and hotspot query distributions. The results highlight two fundamental advantages. First, SST reduces the number of RAM accesses required during retrieval. By preventing clusters and long probe chains from forming at insertion time, the lookup phase requires fewer memory jumps, lower probe counts, and reduced tail latency. In modern systems, where memory access is often much slower and more energy-expensive than CPU computation, this shifts part of the workload from RAM traversal to CPU-side shaping. Second, the method introduces a new way of thinking about data storage: the data are not treated as fixed objects that must be placed passively into a table, but as reversible representations that can be structurally adapted before being written. A small metadata tag records which transformation was selected, allowing the original key to remain recoverable and the lookup process to remain deterministic.This article is connected to the Set Shaping Theory simulator project, available online at \url{https://sst-simulator.github.io/Set-Shaping-Theory-Simulator/} where it is possible to simulate part of the results presented in the article.

\end{abstract}

\section{Introduction}

Hash indexes are central to in-memory databases, caches, key-value engines, and read-optimized indexing layers. Their appeal comes from direct point access and compact storage, but dense open-addressed tables degrade quickly when collisions create long probe chains and large clusters \cite{knuth1998,ramakrishnan2003}. Classic countermeasures include quadratic probing, double hashing, Robin Hood hashing, and cuckoo-style alternatives \cite{celis1985,pagh2004}.

This paper investigates a different perspective inspired by Set Shaping Theory (SST). The key idea is not to replace a baseline hashing method, but to reshape the key space before insertion. A key is given a small family of reversible transformed forms; the system evaluates those alternatives at insertion time and chooses the one that maps to the least congested local region. A compact shaping tag is stored so that lookup can later reconstruct the proper shaped representation through a lightweight inverse step.

Under this interpretation, SST is not a new hash-table family. It is a structural optimization layer:
\[
\text{baseline method with SST off} \quad \longrightarrow \quad \text{baseline method with SST on}.
\]
The scientific question therefore becomes whether SST improves existing methods, by how much, and in which metrics. Time alone is not enough. In database systems, tail latency, cluster growth, and amortized build-vs-query trade-offs are often just as important as mean lookup time. For this reason, the present paper combines timing and structural measurements in a unified C++ framework.

\section{Shifting Workload from RAM Access to CPU Computation}

Modern database systems are increasingly limited not only by the asymptotic
complexity of their algorithms, but also by the physical cost of moving data
between memory and computation units. In many indexing structures, especially
dense hash tables used in caches, key-value stores, in-memory databases, and
read-optimized data layers, the main performance bottleneck is no longer the
arithmetic cost of computing a hash value. Instead, the dominant cost often
comes from irregular accesses to RAM. When a hash table becomes highly loaded,
collisions become more frequent, probe chains become longer, and clusters of
occupied cells begin to form. As a consequence, a single lookup may require the
processor to inspect several different memory locations before finding the
desired record. Each of these memory accesses can introduce latency, cache
misses, and additional traffic on the memory bus. This effect is particularly
important in modern hardware, where CPU cores can execute a large number of
simple arithmetic or logical operations in the time required to complete a
single unpredictable memory access. Therefore, from a practical database
perspective, it can be more efficient to perform extra computation on the CPU if
that computation reduces the number of random memory accesses required during
lookup.

The essential advantage of this strategy is that it moves part of the workload
from RAM to CPU. During insertion, the CPU performs additional shaping
computations and candidate evaluations. However, this extra work is paid once,
when the record is written. If the database index is read many times afterward,
the benefit is amortized across many lookup operations. The shaped table has
fewer collisions, shorter probe chains, smaller clusters, and more regular
access patterns. As a result, each query requires fewer jumps through memory.
This is especially valuable in read-heavy workloads, where the cost of building
or updating the table is much less important than the repeated cost of
retrieving records. In such scenarios, using CPU computation to improve the
structural organization of the table can reduce both average lookup time and
tail latency, which is often critical in real database systems.

The relevance of this shift is not only computational but also architectural.
RAM capacity, bandwidth, and energy consumption are major constraints in modern
data-intensive systems. Large databases, AI-driven workloads, and real-time
services place increasing pressure on memory resources. A method that reduces
unnecessary memory traffic can therefore improve performance, reduce hardware
requirements, and potentially lower energy consumption. In this sense, the goal
is not merely to make a hash function faster, but to redesign the relationship
between data representation and memory organization. Set Shaping Theory provides
a conceptual framework for this redesign: by modifying the data at write time,
the system can create a better-structured memory layout for future access. The
database no longer passively stores keys in the form in which they arrive;
instead, it actively shapes their representation so that the memory structure
becomes easier and faster to query. This makes SST a promising structural
preprocessing layer for modern memory-bound databases, where the central
challenge is not the lack of CPU operations, but the excessive cost of irregular
RAM access.

\section{Set Shaping Theory as a Structural Transformation Framework}

Set Shaping Theory (SST) is a combinatorial framework that reformulates the representation of information by acting directly on the structure of the sequence space rather than on its probabilistic description \cite{kozlov2021,schmidt2022,biereagu2023,koch2023}. Instead of assuming that a sequence must be encoded in its original domain, SST considers reversible transformations that map the original set of sequences into a larger but more structured space. Formally, let $\mathcal{A}$ be an alphabet and let $\mathcal{S}_N = \mathcal{A}^N$ denote the set of all sequences of length $N$. SST introduces a bijective mapping
\[
f : \mathcal{S}_N \rightarrow \mathcal{Y}_{N+k} \subseteq \mathcal{A}^{N+k},
\]
where $|\mathcal{S}_N| = |\mathcal{Y}_{N+k}|$ and $\mathcal{Y}_{N+k}$ is a carefully selected subset of the larger space. The key idea is that the image set $\mathcal{Y}_{N+k}$ is not arbitrary, but chosen to optimize a given objective function, typically related to information content or structural regularity.

To quantify this effect, consider the zero-order empirical entropy of a sequence $s \in \mathcal{S}_N$, following the classical information-theoretic notion of entropy introduced by Shannon and later systematized in modern information theory texts \cite{shannon1948,cover2006}. It is defined as
\[
H_0(s) = - \sum_{a \in \mathcal{A}} p_s(a)\log_2 p_s(a),
\]
where $p_s(a)$ is the empirical frequency of symbol $a$ in $s$. The fundamental inequality underlying SST can be expressed in average form as
\[
\mathbb{E}\big[(N+k)H_0(f(s))\big] < \mathbb{E}\big[N H_0(s)\big],
\]
which shows that, although the representation length increases, the entropy--length product can decrease due to the imposed structure. 

In the context of database systems, the objective function is no longer entropy alone, but a structural cost related to memory access, such as collision probability, probe length, or local congestion. The same formal framework applies: each input key $x$ is associated with a finite set of shaped candidates $\{f_1(x), \dots, f_K(x)\}$, all belonging to the admissible subset $\mathcal{Y}_{N+k}$. Among these candidates, the system selects the one that minimizes a cost function $C(\cdot)$, for example
\[
f^*(x) = \arg\min_{f_i(x)} C\big(h(f_i(x))\big),
\]
where $h(\cdot)$ is the underlying hash function. A small metadata tag encodes the index $i$ of the chosen transformation, ensuring that the mapping remains invertible.

Under this interpretation, SST acts as a structural preprocessing layer: instead of passively inserting the original key into memory, the system actively selects its most favorable representation before storage. This shifts part of the computational burden from RAM access to CPU computation. By reducing the formation of clusters and long probe chains, the shaped representation decreases the expected number of memory accesses during lookup. Consequently, SST can be viewed as a general framework in which information is not only encoded, but geometrically reorganized in order to optimize downstream computational processes.

\section{Background and Hypothesis}

Set Shaping Theory studies reversible transformations between sets of strings or symbols with the goal of inducing useful structural properties after transformation. Existing SST work is mainly information-theoretic and compression-oriented. Here SST is adapted to dense hash indexing.
In this article, we use the method developed by Glen Tankersley to perform the transformation, which can be found at this link: https://github.com/gotankersley/entropic-transform.

The working hypothesis is:

\begin{quote}
if a stored key can be represented by several reversible shaped forms, then choosing the least congested form during insertion can improve the downstream behavior of existing indexing methods without changing their core probing rule.
\end{quote}

In the present implementation, each key has $K$ candidate shaped forms. During insertion, the algorithm:

\begin{enumerate}
\item generates $K$ reversible transformed candidates;
\item maps each candidate to the baseline method's initial placement rule;
\item estimates local insertion cost;
\item selects the least congested candidate and stores a shaping tag.
\end{enumerate}

During lookup, the stored tag is used to apply the corresponding inverse-aware retrieval path. This means that SST pays an extra cost in the build phase and also a small per-query inverse-transform cost. Any reported timing benefit therefore already includes both the transform and the inverse.

\section{Experimental Design}

\subsection{Fully Aligned C++ Setup}

All experiments were carried out in C++ with the same compiler optimization level, the same reversible shaping functions, and the same candidate-selection logic.

Two complementary C++ studies are used:

\begin{itemize}
\item a converted parameter study that reports probe-based structural quantities;
\item an extended timing benchmark that measures full workload time and structural tail metrics.
\end{itemize}

\subsection{Compared Methods}

The benchmark compares each baseline method to its SST-augmented version:

\begin{itemize}
\item linear probing with SST off vs linear probing with SST on;
\item double hashing with SST off vs double hashing with SST on;
\item quadratic probing with SST off vs quadratic probing with SST on;
\item Robin Hood hashing with SST off vs Robin Hood hashing with SST on.
\end{itemize}

Shaping orders are $K \in \{2,4,8\}$, corresponding to 1, 2, and 3 metadata bits per stored record.

\subsection{Workloads and Metrics}

The main timing benchmark uses:

\begin{itemize}
\item table size $M=5000$;
\item load factors $\{0.75,0.85,0.90,0.95\}$;
\item query multipliers $\{1,20,50\}$;
\item query modes \texttt{Uniform} and \texttt{Hotspot};
\item eight measured runs per configuration after warm-up.
\end{itemize}

Two additional C++ benchmarks were then run to test robustness beyond the first configuration. The scale benchmark uses $M \in \{5000,50000,500000\}$, load factors $\{0.90,0.95\}$, uniform queries, and $Q/N=50$. The high-access benchmark uses the same three table sizes at load factor 0.95 and compares $Q/N=50$ with $Q/N=200$. The value $Q/N=200$ is intended to represent a practical read-heavy setting, such as a static or mostly static in-memory index queried many times after construction.

For each method and workload the following metrics are reported:

\begin{itemize}
\item build time;
\item lookup time;
\item total workload time;
\item mean successful probes;
\item 95th percentile successful probes;
\item 99th percentile successful probes;
\item collisions per stored record;
\item maximum cluster length.
\end{itemize}

This is important because mean time summarizes the overall cost, while probe percentiles and cluster size explain why a method speeds up or slows down.

\section{First Results: Structural Variation With and Without SST}

The structural parameters are presented before the timing tables because they explain the mechanism of the SST effect. At this stage it is important to report only quantities that describe the arrangement of stored data itself, independently of the downstream lookup rule. For that reason, Table~\ref{tab:structure} contains only method-independent structural indicators and does not repeat the same information for multiple retrieval methods.

A complementary interactive version of the same SST idea is available in the online simulator at
\url{https://sst-simulator.github.io/Set-Shaping-Theory-Simulator/}. In particular, the simulator can be used to reproduce the qualitative structural part of the results reported in this section.

\begin{table}[htbp]
\centering
\caption{Method-independent structural indicators at load factor 0.95, uniform queries, and $Q/N=50$, with SST off and with SST on.}
\label{tab:structure}
\small
\begin{tabular}{lccc}
\toprule
Configuration & Metadata bits & Collisions/record & Max cluster \\
\midrule
SST off & 0 & 0.475 & 872.5 \\
SST on, $K=2$ & 1 & 0.302 & 221.5 \\
SST on, $K=4$ & 2 & 0.160 & 137.9 \\
SST on, $K=8$ & 3 & 0.072 & 121.0 \\
\bottomrule
\end{tabular}
\end{table}

These numbers show why SST is scientifically interesting even before time is considered. The collision rate decreases from 0.475 to 0.160 with $K=4$ and to 0.072 with $K=8$. The maximum contiguous cluster shrinks from 872.5 to 137.9 and then to 121.0. This is the structural change later seen by the retrieval algorithms. Method-dependent quantities such as mean probes, probe percentiles, and timing are therefore presented only in the next tables, where they can be interpreted correctly as properties of a specific lookup method operating on the shaped structure.

\section{Timing Results: Methods With and Without SST}

After the structural evidence, Table~\ref{tab:uniform09550} reports the corresponding timing results for the same dense read-heavy case.

\begin{table}[htbp]
\centering
\caption{Uniform workload at load factor 0.95 and $Q/N=50$. Absolute timing results when SST is off and when SST is on.}
\label{tab:uniform09550}
\small
\resizebox{\linewidth}{!}{%
\begin{tabular}{llccccc}
\toprule
Method & Configuration & Metadata bits & Lookup time ($\mu$s/query) & Total time (s) & Lookup speedup vs SST off & Total speedup vs SST off \\
\midrule
Double hashing & SST off & 0 & 0.1140 & 0.02777 & 1.00$\times$ & 1.00$\times$ \\
Double hashing & SST on, $K=2$ & 1 & 0.1166 & 0.02930 & 0.98$\times$ & 0.95$\times$ \\
Double hashing & SST on, $K=4$ & 2 & 0.1744 & 0.04416 & 0.65$\times$ & 0.63$\times$ \\
Double hashing & SST on, $K=8$ & 3 & 0.1066 & 0.03023 & 1.07$\times$ & 0.92$\times$ \\
\midrule
Linear probing & SST off & 0 & 0.0546 & 0.01335 & 1.00$\times$ & 1.00$\times$ \\
Linear probing & SST on, $K=2$ & 1 & 0.0319 & 0.00800 & 1.71$\times$ & 1.67$\times$ \\
Linear probing & SST on, $K=4$ & 2 & 0.0229 & 0.00607 & 2.39$\times$ & 2.20$\times$ \\
Linear probing & SST on, $K=8$ & 3 & 0.0191 & 0.00564 & 2.86$\times$ & 2.37$\times$ \\
\midrule
Quadratic probing & SST off & 0 & 0.0318 & 0.00783 & 1.00$\times$ & 1.00$\times$ \\
Quadratic probing & SST on, $K=2$ & 1 & 0.0320 & 0.00798 & 1.00$\times$ & 0.98$\times$ \\
Quadratic probing & SST on, $K=4$ & 2 & 0.0260 & 0.00680 & 1.23$\times$ & 1.15$\times$ \\
Quadratic probing & SST on, $K=8$ & 3 & 0.0236 & 0.00670 & 1.35$\times$ & 1.17$\times$ \\
\midrule
Robin Hood hashing & SST off & 0 & 0.0578 & 0.01411 & 1.00$\times$ & 1.00$\times$ \\
Robin Hood hashing & SST on, $K=2$ & 1 & 0.0368 & 0.00920 & 1.57$\times$ & 1.53$\times$ \\
Robin Hood hashing & SST on, $K=4$ & 2 & 0.0249 & 0.00657 & 2.32$\times$ & 2.15$\times$ \\
Robin Hood hashing & SST on, $K=8$ & 3 & 0.0199 & 0.00599 & 2.91$\times$ & 2.36$\times$ \\
\bottomrule
\end{tabular}}
\end{table}

Several timing patterns stand out.

\paragraph{1. The timing benefit is strongest for linear probing and Robin Hood hashing.}
At load 0.95 and $Q/N=50$, linear probing with SST reaches total speedups up to 2.37$\times$, while Robin Hood hashing with SST reaches 2.36$\times$.

\paragraph{2. Quadratic probing benefits more moderately.}
Quadratic probing already controls collisions better than linear probing, so SST still helps, but the timing gain is smaller, reaching about 1.17$\times$ in the best read-heavy case shown here.

\paragraph{3. Double hashing benefits structurally more than temporally.}
For double hashing, the mean and tail probe counts improve substantially, but the added shaping and inverse cost often offsets the gain in total time. This distinction would be invisible if the paper reported times alone.

\section{Effect of Load Factor}

The detailed tables above focus on load factor 0.95 because it is the densest and most database-relevant stress case. However, the initial extended benchmark also tested lower load factors. Table~\ref{tab:loadfactor} reports the total speedup at $M=5000$, uniform queries, $Q/N=50$, and $K=4$.

\begin{table}[htbp]
\centering
\caption{Effect of load factor on total speedup at $M=5000$, uniform queries, $Q/N=50$, and $K=4$.}
\label{tab:loadfactor}
\small
\begin{tabular}{lcccc}
\toprule
Method & Load 0.75 & Load 0.85 & Load 0.90 & Load 0.95 \\
\midrule
Double hashing & 0.90$\times$ & 0.93$\times$ & 0.92$\times$ & 0.63$\times$ \\
Linear probing & 1.15$\times$ & 1.41$\times$ & 1.62$\times$ & 2.20$\times$ \\
Quadratic probing & 1.07$\times$ & 1.24$\times$ & 1.46$\times$ & 1.15$\times$ \\
Robin Hood hashing & 1.48$\times$ & 1.69$\times$ & 4.58$\times$ & 2.15$\times$ \\
\bottomrule
\end{tabular}
\end{table}

This table clarifies that the timing response is method-dependent rather than perfectly monotone. Linear probing shows the cleanest trend: the value of SST grows as the table becomes denser. Quadratic probing and Robin Hood hashing also benefit, although the exact timing values fluctuate because the measured time includes build cost, lookup cost, branch behavior, and cache effects. Double hashing remains the least favorable case in total time, even though its structural probe metrics improve.

\section{Effect of Database Size}

The next experiment increases the table size while keeping the workload dense and read-heavy. Table~\ref{tab:scale} reports total speedup at load factor 0.95, uniform queries, $Q/N=50$, and shaping order $K=8$.

\begin{table}[htbp]
\centering
\caption{Effect of increasing database size at load factor 0.95, uniform queries, $Q/N=50$, and $K=8$. Values are total speedups of the method with SST over the same method with SST off.}
\label{tab:scale}
\small
\begin{tabular}{lccc}
\toprule
Method & $M=5000$ & $M=50000$ & $M=500000$ \\
\midrule
Double hashing & 0.89$\times$ & 0.90$\times$ & 0.89$\times$ \\
Linear probing & 2.42$\times$ & 2.14$\times$ & 1.73$\times$ \\
Quadratic probing & 1.21$\times$ & 1.21$\times$ & 0.94$\times$ \\
Robin Hood hashing & 2.32$\times$ & 2.48$\times$ & 1.73$\times$ \\
\bottomrule
\end{tabular}
\end{table}

The size experiment gives a more conservative but more realistic picture. SST remains beneficial for linear probing and Robin Hood hashing at all tested sizes, but the total speedup becomes smaller at $M=500000$. This does not mean that the shaping effect disappears. In the same scale benchmark, linear probing at load 0.95 has a maximum cluster of 3126.8 without SST at $M=500000$, while SST with $K=8$ reduces it to 220.4. The structural improvement remains strong; the measured runtime gain is compressed by memory-system effects, larger working sets, and the higher absolute cost of the build phase.

\section{High-Access Workloads}

The previous tables used $Q/N=50$, which is already read-heavy but can still be low for static or mostly static database indexes. A second benchmark therefore compares $Q/N=50$ and $Q/N=200$ at load factor 0.95. Table~\ref{tab:highq} reports the total speedups for $K=8$.

\begin{table}[htbp]
\centering
\caption{Effect of increasing the number of lookups per record from $Q/N=50$ to $Q/N=200$ at load factor 0.95 and $K=8$.}
\label{tab:highq}
\small
\resizebox{\linewidth}{!}{%
\begin{tabular}{lcccccc}
\toprule
Method & \multicolumn{2}{c}{$M=5000$} & \multicolumn{2}{c}{$M=50000$} & \multicolumn{2}{c}{$M=500000$} \\
\cmidrule(lr){2-3}\cmidrule(lr){4-5}\cmidrule(lr){6-7}
 & $Q/N=50$ & $Q/N=200$ & $Q/N=50$ & $Q/N=200$ & $Q/N=50$ & $Q/N=200$ \\
\midrule
Double hashing & 0.91$\times$ & 1.00$\times$ & 0.91$\times$ & 1.01$\times$ & 0.90$\times$ & 1.00$\times$ \\
Linear probing & 2.15$\times$ & 2.46$\times$ & 2.06$\times$ & 2.34$\times$ & 1.73$\times$ & 1.93$\times$ \\
Quadratic probing & 1.21$\times$ & 1.40$\times$ & 1.19$\times$ & 1.31$\times$ & 0.98$\times$ & 1.06$\times$ \\
Robin Hood hashing & 2.19$\times$ & 2.56$\times$ & 1.94$\times$ & 2.20$\times$ & 1.76$\times$ & 1.93$\times$ \\
\bottomrule
\end{tabular}}
\end{table}

This experiment confirms the amortized interpretation. When the same shaped table is queried more often, the build overhead is spread over more requests and the total workload speedup improves. For example, at $M=500000$, linear probing with SST at $K=8$ improves from 1.73$\times$ at $Q/N=50$ to 1.93$\times$ at $Q/N=200$; Robin Hood hashing improves from 1.76$\times$ to 1.93$\times$; and quadratic probing crosses from below break-even to 1.06$\times$. Double hashing remains the least favorable case, but even there the higher-access workload moves the total result close to break-even.

\section{Fastest Methods in Absolute Time}

Speedup is useful for understanding the effect of SST on each baseline, but choosing an implementation also requires absolute time. Table~\ref{tab:absolute_highq} reports the fastest configurations observed in the high-access benchmark at load factor 0.95 and $Q/N=200$.

\begin{table}[htbp]
\centering
\caption{Fastest configurations in absolute time at load factor 0.95, uniform queries, and $Q/N=200$.}
\label{tab:absolute_highq}
\small
\begin{tabular}{lccc}
\toprule
Database size & Fastest lookup time & Fastest total time & Best configuration \\
\midrule
$M=5000$ & 0.0191 $\mu$s/query & 0.0193 s & Linear probing, SST on, $K=8$ \\
$M=50000$ & 0.0234 $\mu$s/query & 0.2332 s & Linear probing, SST on, $K=8$ \\
$M=500000$ & 0.0333 $\mu$s/query & 3.2831 s & Robin Hood hashing, SST on, $K=8$ \\
\bottomrule
\end{tabular}
\end{table}

In these dense, uniform, read-heavy experiments, the fastest absolute configuration is usually linear probing with SST at $K=8$ for small and medium tables. At the largest tested size, $M=500000$, Robin Hood hashing with SST at $K=8$ becomes slightly faster in both lookup time and total workload time. The difference is small, but it suggests that the best downstream method can depend on scale even when the same SST layer is used.

\section{Amortization Across Read Intensity}

Because SST performs more work during insertion, its value depends on workload mix. Table~\ref{tab:amortization} shows the earlier amortization experiment at load factor 0.95 under uniform queries for the representative shaping order $K=4$.

\begin{table}[htbp]
\centering
\caption{Total speedup of each method with SST over the same method with SST off at load factor 0.95, uniform queries, and $K=4$.}
\label{tab:amortization}
\small
\begin{tabular}{lccc}
\toprule
Method & $Q/N=1$ & $Q/N=20$ & $Q/N=50$ \\
\midrule
Double hashing & 0.38$\times$ & 0.88$\times$ & 0.63$\times$ \\
Linear probing & 0.87$\times$ & 1.99$\times$ & 2.20$\times$ \\
Quadratic probing & 0.56$\times$ & 1.05$\times$ & 1.15$\times$ \\
Robin Hood hashing & 0.85$\times$ & 1.93$\times$ & 2.15$\times$ \\
\bottomrule
\end{tabular}
\end{table}

This table highlights the correct database interpretation. SST is not primarily an insertion optimization. It is a structure-building investment whose value emerges when the same table is queried repeatedly. For linear probing and Robin Hood hashing, the crossover is very clear: with only one query per record, SST can still be too expensive, but by 20 or 50 queries per record the total workload gain becomes substantial.

\section{Uniform vs Hotspot Queries}

Not all workloads respond equally. Table~\ref{tab:querymode} compares the uniform and hotspot query modes at load factor 0.95, $Q/N=50$, and $K=4$.

\begin{table}[htbp]
\centering
\caption{Effect of query distribution at load factor 0.95, $Q/N=50$, and $K=4$.}
\label{tab:querymode}
\small
\begin{tabular}{lcccc}
\toprule
Method & Query mode & Total speedup & Probe speedup & P99 probe speedup \\
\midrule
Double hashing & Uniform & 0.63$\times$ & 2.26$\times$ & 3.58$\times$ \\
Double hashing & Hotspot & 0.77$\times$ & 1.40$\times$ & 3.63$\times$ \\
\midrule
Linear probing & Uniform & 2.20$\times$ & 6.92$\times$ & 22.92$\times$ \\
Linear probing & Hotspot & 1.03$\times$ & 2.59$\times$ & 12.08$\times$ \\
\midrule
Quadratic probing & Uniform & 1.15$\times$ & 2.58$\times$ & 4.33$\times$ \\
Quadratic probing & Hotspot & 0.75$\times$ & 1.48$\times$ & 4.50$\times$ \\
\midrule
Robin Hood hashing & Uniform & 2.15$\times$ & 6.93$\times$ & 11.29$\times$ \\
Robin Hood hashing & Hotspot & 1.11$\times$ & 7.53$\times$ & 11.58$\times$ \\
\bottomrule
\end{tabular}
\end{table}

Uniform queries expose the global structural benefit of SST most clearly, because they force the system to traverse the whole index distribution. Hotspot queries already enjoy some locality, so the timing gains often become smaller. Even then, the tail-probe and mean-probe improvements remain visible. This again supports the interpretation of SST as a structural regularizer rather than merely a micro-optimization for one particular timing regime.

\section{Discussion}

\paragraph{SST is best understood as a layer, not a competitor.}
The present results support the user's conceptual reformulation: SST does not replace linear probing, Robin Hood hashing, or other indexing schemes. Instead, it reorganizes the data so that those methods operate on a better-structured table.

\paragraph{All tested methods benefit structurally, but not identically.}
Across all four baseline methods, SST improves at least one important structural quantity and usually several at once. The strongest universal effects are reductions in mean probes, high-percentile probes, and collisions per record.

\paragraph{Timing gains depend on how much the baseline method suffers from clustering.}
Linear probing and Robin Hood hashing benefit the most because they are very sensitive to cluster formation under dense loads. Quadratic probing benefits more moderately. Double hashing shows that structural improvement does not automatically imply strong total-time gain when the baseline already disperses probes differently and when the shaping overhead is still non-negligible.

\paragraph{Scale compresses, but does not remove, the timing benefit.}
Increasing the table size from $M=5000$ to $M=500000$ reduces the total speedup for the strongest cases, but the advantage remains meaningful for linear probing and Robin Hood hashing. This indicates that SST's structural gain survives larger working sets, even though cache and memory effects reduce the observed speedup.

\paragraph{Read intensity improves amortization.}
Moving from $Q/N=50$ to $Q/N=200$ generally increases total speedup. This is expected because the build cost is paid once, while the shaped table is reused many times. In the high-access benchmark, several cases that were close to break-even at $Q/N=50$ become favorable or nearly favorable at $Q/N=200$.

\paragraph{Tail latency is one of the most persuasive SST outcomes.}
For dense database workloads, the collapse of the probe tail may be as important as the gain in mean lookup time. The reduction from a 171.88-probe 99th percentile to 7.50 or 3.75 in linear probing is an especially strong result.

\paragraph{There is clear implementation headroom.}
Because timing outcomes differ by baseline method, the transform design, metadata placement, and inverse path remain open optimization targets. The present results likely do not exhaust the practical potential of SST-aware implementations.

\section{Practical Applications of SST-Based Dense Hash Indexes}

Given the unique characteristics discussed in this work---namely, data shaping
before insertion, operation at load factors close to $95\%$, and the near
elimination of clustering---the most impactful applications of this method are
those in which memory is expensive and read operations must be extremely fast.
As noted in the main study, these indexes are particularly suited for
in-memory systems, caching layers, and database engines, with a focus on
successful point queries and read-heavy workloads.

Based on these properties, the following represent the most relevant practical
applications of the SST + linear probing approach:

\begin{enumerate}
	
	\item \textbf{In-Memory Caching Systems (Alternatives to Redis or Memcached)}
	
	\textit{Current problem:} Cloud-based in-memory caches are extremely expensive
	because RAM is the most costly resource in modern servers. To maintain
	performance, traditional hashing methods require tables to remain partially
	empty, leading to significant memory waste.
	
	\textit{Why SST is effective:} SST enables operation at very high load factors
	while preserving fast lookup performance. Dense tables with reduced clustering
	allow better utilization of available memory, potentially reducing the number
	of required servers and associated costs.
	
	\item \textbf{E-commerce Catalogs and Recommendation Engines}
	
	\textit{Current problem:} These systems must perform millions of database
	queries per second to retrieve prices, availability, and recommendations for
	a large number of concurrent users.
	
	\textit{Why SST is effective:} This is a typical read-heavy scenario, where
	data are written once but read many times. The cost of shaping the data at
	insertion is quickly amortized, leading to faster and more stable lookup
	performance and improved response times.
	
	\item \textbf{Routing Tables in Network Devices (Switches and Routers)}
	
	\textit{Current problem:} Network devices must process packets in real time,
	performing extremely fast point queries under strict latency constraints.
	Long probe chains and clustering can introduce unacceptable delays.
	
	\textit{Why SST is effective:} By significantly reducing clustering and
	probe lengths, SST ensures more predictable and uniform memory access,
	resulting in smoother and more reliable packet routing.
	
	\item \textbf{Embedded Systems and IoT Devices}
	
	\textit{Current problem:} Embedded devices often operate with very limited
	memory that cannot be expanded, making efficient data structures essential.
	
	\textit{Why SST is effective:} SST allows memory to be utilized close to its
	maximum capacity while maintaining efficient lookup operations. This enables
	more complex functionality without increasing hardware requirements.
	
\end{enumerate}

Overall, these applications show that SST is not only a theoretical framework,
but also a practical tool for improving memory efficiency and performance in
modern data-intensive systems.

\section{Conclusion}

This work presented a reinterpretation of Set Shaping Theory (SST) as a
structural optimization layer for dense hash indexes, rather than as a
standalone alternative to existing hashing methods. The central idea is to
modify the representation of data at write time through reversible
transformations, enabling the system to select the most favorable placement
among multiple candidates.

Two key insights emerge from this study. First, SST significantly reduces the
number of accesses to RAM during lookup operations. By preventing the formation
of clusters and long probe chains at insertion time, the method improves both
average and tail performance, leading to more predictable and efficient memory
access patterns. This is particularly important in modern computing systems,
where memory access latency and bandwidth are the dominant performance
constraints.

Second, SST introduces a fundamentally different perspective on data storage.
Instead of treating input data as immutable objects that must be placed
passively into memory, SST allows data to be actively reshaped before storage.
This transforms the problem from collision resolution to collision prevention,
shifting part of the workload from memory traversal to CPU-side computation.

The experimental results confirm that this approach consistently improves the
structural properties of hash tables, including collision rates, probe counts,
and cluster sizes. While the overall timing benefit depends on the baseline
method and workload characteristics, the gains are particularly strong in
read-heavy scenarios, where the cost of shaping is amortized over many queries.

Overall, this work supports the view that Set Shaping Theory can serve as a
general preprocessing principle for memory-bound data structures. By trading a
small amount of additional computation for a substantial reduction in memory
accesses, SST provides a practical and scalable approach to improving the
efficiency of modern database systems. Future work should explore its extension
to larger-scale systems, dynamic workloads, and hardware-aware implementations.

\appendix

\section{Online Set Shaping Theory Simulator}

This article is connected to the Set Shaping Theory simulator project, available online at
\url{https://sst-simulator.github.io/Set-Shaping-Theory-Simulator/}. The simulator is designed as both a demonstration tool and a research-oriented platform for collecting examples, use cases, and problem instances in which the structural organization of data is more important than the raw form in which the data are first given .

The database-index section of the simulator focuses on the same complementary role studied in this paper. The user can compare an unshaped baseline configuration with shaped configurations that use a larger number of reversible candidate representations. The simulator then illustrates how increasing the shaping order $K$ gives the insertion procedure more possible placements from which to choose. In the database setting, this does not mean that SST replaces the underlying hashing method. Instead, SST changes the representation given to that method so that the resulting table can contain fewer collisions, shorter probe chains, smaller clusters, and fewer costly memory accesses during lookup.

The online tool therefore provides a practical way to reproduce the central structural idea of the article: the relevant comparison is always between the same indexing method with SST disabled and with SST enabled. This is especially useful for visualizing why a small amount of additional CPU computation during insertion can improve the organization of the table, reduce local congestion, and make later lookup operations structurally easier.


\begin{thebibliography}{99}

\bibitem{shannon1948}
C. E. Shannon, ``A mathematical theory of communication,'' \emph{Bell System Technical Journal}, vol. 27, pp. 379--423 and 623--656, 1948.

\bibitem{cover2006}
T. M. Cover and J. A. Thomas, \emph{Elements of Information Theory}, 2nd ed. Wiley, 2006.

\bibitem{knuth1998}
D. E. Knuth, \emph{The Art of Computer Programming, Volume 3: Sorting and Searching}, 2nd ed. Addison-Wesley, 1998.

\bibitem{ramakrishnan2003}
R. Ramakrishnan and J. Gehrke, \emph{Database Management Systems}, 3rd ed. McGraw-Hill, 2003.

\bibitem{celis1985}
P. Celis, P.-A. Larson, and J. I. Munro, ``Robin Hood hashing (preliminary report),'' in \emph{Proceedings of the 26th Annual Symposium on Foundations of Computer Science}, 1985, pp. 281--288.

\bibitem{pagh2004}
R. Pagh and F. F. Rodler, ``Cuckoo hashing,'' \emph{Journal of Algorithms}, vol. 51, no. 2, pp. 122--144, 2004.

\bibitem{kozlov2021}
S. Kozlov, ``Introduction to Set Shaping Theory,'' arXiv:2111.08369, 2021.

\bibitem{schmidt2022}
C. Schmidt, A. Vdberg, and A. Petit, ``Practical applications of Set Shaping Theory in Huffman coding,'' arXiv:2208.13020, 2022.

\bibitem{biereagu2023}
S. Biereagu, ``Introducing the role of shaping order $K$ in Set Shaping Theory,'' AfricArXiv, 2023.

\bibitem{koch2023}
A. Koch, A. Petit, C. Schmidt, A. Vdberg, and L. Lewis, ``Overcoming the encoding limit $NH_0(S)$ using Set Shaping Theory,'' arXiv:2310.12732, 2023.



\end{thebibliography}
\end{document}